\documentclass[runningheads]{llncs}
\usepackage{amsmath,graphicx}

\usepackage[utf8]{inputenc} 
\usepackage[T1]{fontenc}    
\usepackage{hyperref}       
\usepackage{url}            
\usepackage{booktabs}       
\usepackage{amsfonts}       
\usepackage{nicefrac}       
\usepackage{microtype}      
\usepackage{xcolor}         
\usepackage{amssymb}
\DeclareMathAlphabet\mathbfcal{OMS}{cmsy}{b}{n}

\usepackage{xurl}

\usepackage{algorithm}
\usepackage{algpseudocode}

\newcommand{\rvx}[1]{\boldsymbol{#1}}
\newcommand{\rulesep}{\unskip\ \vrule\ }

\begin{document}
\title{Geometry-preserving Lie group integrators for differential equations on the manifold of symmetric positive definite matrices\thanks{This work was supported by Agence Nationale de la Recherche under grant ANR-21-CE48-0005 LEMONADE.}}
\titlerunning{Lie group integrators for ODEs on Symmetric Positive Definite matrices}
%

\author{Lucas Drumetz\inst{1} \and
Alexandre Reiffers-Masson\inst{1} \and
Naoufal El Bekri\inst{1,2} \and Franck Vermet\inst{2}}

\authorrunning{L. Drumetz et al.}
%
\institute{IMT Atlantique, Lab-STICC, UMR CNRS 6285, Brest, France \\
\and
Univ Brest,  UMR CNRS 6205, Laboratoire de Mathématiques de Bretagne Atlantique, France \\
\email{lucas.drumetz@imt-atlantique.fr}}
\maketitle              
\begin{abstract}
In many applications, one encounters time series that lie on manifolds rather than a Euclidean space. In particular, covariance matrices are ubiquitous mathematical objects that have a non Euclidean structure. The application of Euclidean methods to integrate differential equations lying on such objects does not respect the geometry of the manifold, which can cause many numerical issues. In this paper, we propose to use Lie group methods to define geometry-preserving numerical integration schemes on the manifold of symmetric positive definite matrices. These can be applied to a number of differential equations on covariance matrices of practical interest. We show that they are more stable and robust than other classical or naive integration schemes on an example.

\end{abstract}
\section{Introduction}
\vspace{-0.1cm}

Ordinary Differential Equations (ODEs) arise everywhere in science, and are a fundamental tool to describe continuous-time dynamical systems~\cite{jordan2007nonlinear}. However, numerical integration is almost always required to obtain approximate solutions. Most of the time, the variable to integrate lives in a Euclidean space, typically $\mathbb{R}^n$. One can then choose from many methods, ranging from simple explicit/implicit Euler or Runge-Kutta methods to adaptive time step schemes~\cite{butcher2016numerical}.\\
In a number of situations, one may require that the variable to integrate lies on a manifold \cite{hairer2011solving}. Examples include flows on spheres, rotation or covariance matrices (or other matrix manifolds)\cite{iserles2000lie}... For embedded submanifolds of $\mathbb{R}^{n}$, though the underlying vector space makes it possible to apply classical ODE integration methods, they cannot guarantee that the numerical solution stays on the manifold at each time step. Providing these guarantees is crucial for subsequent uses of the solution, e.g. computing geodesic distances for Riemannian manifolds~\cite{absil2009optimization}, or simply keeping the structural or physical interpretation of a variable. \\
Formally, the flow of a smooth vector field on a smooth manifold $\mathcal{M}$ generated by an ODE with the initial condition $x(0) = x_0$, writes:
\vspace{-0.262cm}
    \begin{equation}
        \frac{dx}{dt} = F|_{x(t)}(t),
        \label{ODE}
    \end{equation}
 where $x\in \mathcal{M}$ and $F|_{x(t)}(t)$ is a (time dependent) tangent vector to  $\mathcal{M}$ at $x$, and $F: [0,+\infty[ \rightarrow \mathfrak{X}(\mathcal{M})$, with $\frak{X}(\mathcal{M})$ the set of smooth vector fields on $\mathcal{M}$ \cite{owren2018lie}. To integrate such ODEs when $\mathcal{M}$ has the additional structure of a Lie group, several frameworks were developed under the umbrella term "Lie group integrators" \cite{celledoni2014introduction}. Interestingly, these methods can be extended to any smooth manifold acted upon transitively by a Lie group~\cite{iserles2000lie}, that is to any homogeneous space.\\
 In this paper, we focus more specifically on the manifold of $n\times n$ symmetric positive definite (SPD) matrices, denoted as $\mathrm{Sym}^{+}_n$~\cite{bhatia2009positive}. It is the manifold of (nondegenerate) covariance matrices, fundamental for multivariate statistics. Flows of covariance matrices arise in many applications, such as Brain Computer Interfaces (BCI) \cite{zanini2017transfer}, Diffusion Tensor Image processing~\cite{dryden2009non}, finance~\cite{dellaportas2004large}, control~\cite{bacsar1998dynamic}, or data assimilation~\cite{evensen2022data}, to represent the evolution of second order moments of random variables. For example, the second order moments of the solution of Stochastic Differential Equations (SDEs) provide simplified and interpretable representations of stochastic processes, though partial in general. In data assimilation, quantifying and propagating the uncertainty of the state variable is crucial and is done in practice using covariance matrices \cite{pannekoucke2016parametric}. Solutions of covariance matrix ODEs which are not SPD are meaningless in terms of statistical interpretation. Thus, we focus on equations similar to~\eqref{ODE}, where the manifold $\mathcal{M}$ is $ \mathrm{Sym}^{+}_n$, and the RHS of~\eqref{ODE} is a symmetric matrix (an element of the tangent space of $\mathrm{Sym}^{+}_n$). In spite of the ubiquiteness of covariance matrices, to the best of our knowledge, Lie group integrators have not been considered yet for $\mathrm{Sym}^{+}_n$.\\
 Our contributions are multiple. i) We show that in the case of $\mathrm{Sym}^{+}_n$, for small enough time steps, classical methods actually remain in the manifold, but, with moderately big time steps, the iterates may cross the boundary of $\mathrm{Sym}^{+}_n$, (consisting in positive semidefinite matrices), leading to meaningless solutions or even diverging algorithms. ii) We propose to use a Lie group action of invertible matrices on  $\mathrm{Sym}^{+}_n$ to turn the latter into a homogeneous space, that can be used to integrate many equations of interest. iii) From there, we design Lie group versions of the Euler and Runge-Kutta 4 (RK4) methods (applicable to many other schemes) on $\mathrm{Sym}^{+}_n$. iv) We conduct experiments an example ODE on $\mathrm{Sym}^{+}_n$ related to a multivariate SDE. They indicate that our integrators perform better than classical or naive schemes, in particular when the integration step is large.
\section{Theoretical results}
In this section, we provide sufficient conditions on the integration time step $\rho$ of the form $\mathbf{P}_{i+1} = \mathbf{P}_{i} + \rho \mathbf{T}$ to either stay in or leave $\mathrm{Sym}^{+}_n$. The results are stated in the following theorem:
\begin{theorem}
Let $\mathbf{P}\in \mathrm{Sym}_{n}^{+}$, and $\mathbf{T}\in \mathrm{Sym}_n$. We denote as $\lambda_1 \leq ... \leq \lambda_n$ the eigenvalues of $\mathbf{P}$ and as $\nu_1 \leq ... \leq \nu_n $ the (real) eigenvalues of $\mathbf{T}$. We define a set $S = \{(i,j) \in \mathbb{N}_*^2,1 \leq i,j \leq n, i+j = n+1, \nu_i \leq 0 \}$. Then the following holds:
\begin{enumerate}
    \item If $\mathbf{T}$ is positive semidefinite, then $\forall \rho \in \mathbb{R}^{+}$, $\mathbf{P}+\rho \mathbf{T} \in \mathrm{Sym}_{n}^{+}$.
    \item When $\mathbf{T}$ has at least one negative eigenvalue, $S$ in nonempty and if
    \begin{equation*}
    \rho \geq \rho_{\textrm{min}} = \min_{(i,j) \in S} -\frac{\lambda_j}{\nu_i}    \end{equation*}
    then $\mathbf{P}+\rho \mathbf{T} \notin \mathrm{Sym}_{n}^{+}$.
    \item When $\mathbf{T}$ has at least one negative eigenvalue, if
    \begin{equation*}
    \rho < \rho_\textrm{max} =  -\frac{\lambda_1}{\nu_1},
    \end{equation*}
    then $\mathbf{P}+\rho \mathbf{T} \in \mathrm{Sym}_{n}^{+}$.
\end{enumerate}
\label{th}
\vspace{-0.1cm}
\end{theorem}
The proof can be found in the appendices. This theorem means that when the vector used for the update happens to be positive semidefinite, then a classical Euler or RK step will remain on $\mathrm{Sym}_{n}^{+}$. However, in the general case, that vector may have negative eigenvalues, and we have shown that the next iterate will leave (resp. remain on) $\mathrm{Sym}_{n}^{+}$ for time steps that are too large (resp. small enough). In practice, the lower bound is obtained by searching the best value in $S$, whose cardinal is the number of negative eigenvalues of $\mathbf{T}$. Time steps whose values lie in between both bounds may or may not stay on the manifold. Thus, algorithms at least  guaranteeing that the trajectory remains on $\mathrm{Sym}_{n}^{+}$ are necessary. We stress that even when iterates of classical methods remain on $\mathrm{Sym}_{n}^{+}$, the geometry of the manifold is not accounted for, which may lead to low quality solutions nonetheless.
\vspace{-0.3cm}
\section{Background on Lie group integrators}
\label{sec:lie}
The general idea behind basic Lie group methods relies on the fact that the flow of a simple class of vector fields on the Lie group is easy to compute via the Lie exponential map. The corresponding equations are analogous to the linear ODE $d \mathbf{x}/dt = \mathbf{A}\mathbf{x}$ in Euclidean spaces. In the case of general ODEs, discretizing by temporarily fixing the vector field of a general equation with a nonconstant "$\mathbf{A}$" (depending on $x$ and $t$), an approximate Euler-like scheme can be computed step by step. Higher order schemes such as RK4 require a vector space structure to be able to manipulate vector fields at different locations and times. Then, we have to further translate the ODE from the Lie group to the Lie algebra.\\
Interestingly, all these methods can be effortlessly extended to smooth manifolds on which we can find a \textit{transitive Lie group action}~\cite{owren2018lie,celledoni2014introduction} (i.e. homogeneous spaces). Throughout this section, we follow~\cite{iserles2000lie} (Chap.2). Here, we limit ourselves to matrix Lie groups for simplicity. Let $\mathcal{M}$ be the smooth manifold, $G$ the matrix Lie group. A smooth map $\Lambda:\ G \times \mathcal{M} \rightarrow \ \mathcal{M}$
is a Lie group action if and only if
\begin{align}
    & \forall x \in \mathcal{M},  \Lambda(\mathbf{I},x) = x , \label{neutral} \\
    & \forall x\in \mathcal{M}, \forall \ \mathbf{A},\mathbf{B} \in G, \Lambda( \mathbf{A},\Lambda( \mathbf{B},x)) = \Lambda( \mathbf{AB},x).
\end{align}
$\Lambda$ is further said to be transitive if 
\begin{equation}
\forall x,y \in \mathcal{M}, \exists  \mathbf{A} \in G, \Lambda( \mathbf{A},x) = y.    
\end{equation}
This means that any point of $\mathcal{M}$ can be reached from any other using the group action with an element of G. 
Equivalently, the group action has only one orbit.\\
To every Lie group $G$ is associated a Lie algebra $\mathfrak{g}$, which is a vector space (the tangent space to the Lie group at the identity). For matrix Lie groups, the Lie algebra is also a set of matrices. Associated to a transitive Lie group action on a smooth manifold is a \emph{algebra action} that translates the group action into an infinitesimal action giving an element of tangent space to the manifold at every point. It determines the type of equations that can be dealt with. It is defined~\cite{iserles2000lie} (Lemma 2.6) as a map $\lambda_{*}:\mathfrak{g} \times \mathcal{M} \rightarrow \frak{X}(\mathcal{M})$ such that, for a given point $x\in \mathcal{M}$:
    \begin{equation}
        \lambda_{*}( \mathbf{A})(x) = \frac{d}{ds} \Lambda(\boldsymbol{\rho}(s),x) |_{s=0},
        \label{homomorphism}
    \end{equation}
where $\boldsymbol{\rho}(s)$ is a smooth curve on $G$, parameterized by a scalar $s$, with initial value $\boldsymbol{\rho}(0) = \mathbf{I}$ and initial speed $ \mathbf{A}\in \mathfrak{g}$ ($\boldsymbol{\rho}'(0) =  \mathbf{A}$).
On matrix groups, $\boldsymbol{\rho}$ can be written as a Taylor expansion:
\begin{equation}
    \boldsymbol{\rho}(s)=\mathbf{I}+ s\mathbf{A} + o(s).
\end{equation}
Intuitively, this curve represents a direction $\mathbf{A}$ on the Lie algebra towards which we can move infinitesimally from any point in $G$.\\
The first step is to write the differential equation in terms of the algebra action associated with an adequately chosen group action ($\Lambda$, with the associated $\lambda_*$):
        \begin{equation}
        \frac{dx}{dt} = F|_{x(t)}(t) = \lambda_{*}(\boldsymbol{\xi}(x(t),t))(x(t)),
        \label{flow}
    \end{equation}
    with the initial condition $x(t_i) = x_{i}$. $\boldsymbol{\xi}:\mathcal{M} \times \mathbb{R}^{+} \rightarrow  \ \mathfrak{g}$ is a smooth function.\\
Thanks to \cite{iserles2000lie} (Lemma 2.7), we know there exists $\mathbf{Y}(t)\in G$ following the differential equation (with $\mathbf{Y}(0) = \mathbf{I}$) :
        \begin{equation}
        \frac{d\mathbf{Y}}{dt} = \boldsymbol{\xi}(x,t) \mathbf{Y}(t).
        \label{Lie_flow}
        \end{equation}
such that $\Lambda(\mathbf{Y}(t),x_0)$ is the solution of Eq~\eqref{flow}. At this step, we can already design a simple Lie-Euler method by temporarily fixing $\boldsymbol{\xi}$ to its current value $\boldsymbol{\xi}(x_i,t_i)$. Then, the solution of the "frozen" ODE~\eqref{Lie_flow} on $G$ is~\cite{iserles2000lie} (Theorem 2.8):
\begin{equation}
    \tilde{\mathbf{Y}}(t) = \exp(t\boldsymbol{\xi}(x_i,t_i)),
    \label{Lieflow}
\end{equation}
with $\exp$ the matrix (Lie group) exponential. By setting $t = t_{i+1} = t_i + h$, we obtain the next iterate on $G$. We come back to $\mathcal{M}$ using the group action: 
\begin{equation}
x_{i+1} = x(t_i + h) = \Lambda(\tilde{\mathbf{Y}}(t + h),x_i).
\label{group_action}
\end{equation}
The Lie-Euler method is summarized in Algorithm~\ref{le}.
\begin{algorithm}
\caption{Lie-Euler algorithm}
\begin{algorithmic}
\Require Transitive group action $\Lambda$ of $G$ on $\mathcal{M}$, associated algebra action $\lambda_*$. Smooth function $\boldsymbol{\xi}$ from $\mathcal{M}$ to $\mathfrak{g}$ such that the ODE on $\mathcal{M}$ writes as in~\ref{flow}. Initial condition $x_0 \in \mathcal{M}$. Step size $h$. Number of time steps $N$.
\State $i = 0$
\State $x_i \gets x_0$
\While{$i \leq N$}
\State$\mathbf{A} \gets h\boldsymbol{\xi}(x_i,t_i)$
\State $x_{i+1} \gets \Lambda(\exp(\boldsymbol{\mathbf{A}}),x_i))$
\State $i \gets i+1$
\EndWhile
\end{algorithmic}
\label{le}
\end{algorithm}
For other schemes requiring to combine the current iterate on the manifold and intermediary values of the flow, e.g. RK4, we cannot form linear combinations of evaluations of the vector field for different times $t$ and locations $x$, since those live on different tangent spaces, nor can we add them to points on $\mathcal{M}$. Thus, we need an ODE defined on a Euclidean space: the Lie algebra $\mathfrak{g}$. Fortunately,~\cite{iserles2000lie} (Lemma 3.1) tells us that the solution of Eq.~\eqref{Lie_flow} writes as:
\begin{equation}
    \mathbf{Y}(t) = \exp(\boldsymbol{\theta}(t))
\end{equation}
where $\boldsymbol{\theta}(t) \in \mathfrak{g}$ solves an ODE in the Lie algebra:
\begin{equation}
    \frac{d\boldsymbol{\theta}}{dt} = \mathrm{dexp}^{-1}_{\boldsymbol{\theta}(t)}(\boldsymbol{\xi}(x,t)) 
    \label{LA_eq}
\end{equation}
with $\boldsymbol{\theta}(0) = \mathbf{0}$. $\mathrm{dexp}^{-1}_{\boldsymbol{\theta}(t)}$ is the inverse of the derivative of the matrix exponential at $\boldsymbol{\theta}(t)$. Its approximation at order 4 (see~\cite{iserles2000lie}, (Def. 2.18)), for any $\mathbf{A}, \boldsymbol{\theta} \in \mathfrak{g}$ is
        \begin{equation}
        d \exp_{\boldsymbol{\theta}}^{-1}(\mathbf{A}) = \mathbf{A} + \frac{1}{2} [\boldsymbol{\theta},\mathbf{A}] + \frac{1}{12} [\boldsymbol{\theta} ,[\boldsymbol{\theta},\mathbf{A}]] 
       - \frac{1}{720} [\boldsymbol{\theta} ,[\boldsymbol{\theta}, [\boldsymbol{\theta} ,[\boldsymbol{\theta},\mathbf{A}]]]] + o(||\boldsymbol{\theta}||^4)
        \label{dexpm}
    \end{equation}
using the matrix commutator $[\mathbf{A},\mathbf{B}] = \mathbf{AB} - \mathbf{BA}$. On Eq.~\eqref{LA_eq}, we can apply a classical RK4 scheme. Each necessary evaluation $\mathbf{K}_i\in \mathfrak{g}$ of the vector field can be done using the algebra action $\lambda_*$ at the right location and time. In the Lie algebra, computing weighted combinations of these evaluations is possible, giving a final vector field $\boldsymbol{\Theta} \in \mathfrak{g}$. Finally, we obtain $x_{i+1}$ on the manifold via:
\begin{equation}
    x_{i+1} = x(t_i +h) =  \Lambda(\exp(\boldsymbol{\Theta}),x_{i}).
\end{equation}
Error order is guaranteed to be the same as the Euclidean scheme as long as there are enough terms (we gave enough for RK4) in the approximation~\eqref{dexpm}. The so-called Runge-Kutta-Munthe-Kaas (RKMS) 4 method is an adaptation of the general RKMS algorithm~\cite{iserles2000lie}, and can be extended to any choice of coefficients defining a classical RK method. This method is summarized in Algorithm~\ref{alg:cap}.
\begin{algorithm}[t]
\caption{Runge-Kutta-Munthe-Kaas 4 algorithm}\label{alg:cap}
\begin{algorithmic}
\Require Transitive group action $\Lambda$ of $G$ on $\mathcal{M}$, associated algebra action $\lambda_*$. Smooth function $\boldsymbol{\xi}$ from $\mathcal{M}$ to $\mathfrak{g}$ such that the ODE on $\mathcal{M}$ writes as in~\ref{homomorphism}. Initial condition $x_0 \in \mathcal{M}$. Step size $h$. Number of time steps $N$.




\State $i = 0$
\State $x_i \gets x_0$
\While{$i \leq N$}
\State$\mathbf{A}_1 \gets h\boldsymbol{\xi}(x_i,t_i)$
\State$\mathbf{K}_1 \gets \mathbf{A}_1$
\State$\mathbf{A}_2 \gets h\boldsymbol{\xi}(\Lambda(\exp(\mathbf{K}_1/2),x_i),t_i+ h/2)$
\State$\mathbf{K}_2 \gets \mathrm{dexp}^{-1}_{\mathbf{K}_1/2}(\mathbf{A}_2)$

\State$\mathbf{A}_3 \gets h\boldsymbol{\xi}(\Lambda(\exp(\mathbf{K}_2/2),x_i),t_i+h/2)$
\State$\mathbf{K}_3 \gets \mathrm{dexp}^{-1}_{\mathbf{K}_2/2}(\mathbf{A}_3)$

\State$\mathbf{A}_4 \gets h\boldsymbol{\xi}(\Lambda(\exp(\mathbf{K}_3),x_i),t_i)$
\State$\mathbf{K}_4 \gets \mathrm{dexp}^{-1}_{\mathbf{K}_3}(\mathbf{A}_4)$

\State $\boldsymbol{\Theta} = \mathbf{K}_1/6 + \mathbf{K}_2/3 + \mathbf{K}_3/3 + \mathbf{K}_4/6$  \Comment{This operation is possible and stable because all summands are in $\mathfrak{g}$}
\State $x_{i+1} \gets \Lambda(\exp(\boldsymbol{\Theta}),x_i))$
\State $i \gets i+1$

\EndWhile
\end{algorithmic}
\label{rkmk}
\end{algorithm}

\section{Application to the SPD manifold}
\label{sec:cov}
In this section, we instantiate the Lie group integrator framework on $\mathrm{Sym}_{n}$ using two different actions by two Lie Groups. The first one, which has a nice interpretation in terms of transformation of covariance matrices by linear functions, is the action of invertible matrices on SPD matrices by congruence. The second is the action of the symplectic group $SP(2n)$ on $\mathrm{Sym}^{+}_n$, seen as the imaginary part of the Siegel upper half-space $SH(n)$. This action and the objects of interest are more involved (actually they generalize the first framework), but this second framework has deep connections with symplectic geometry and equations that arise in linear quadratic optimal control, or Kalman Bucy filtering (in continuous time).

\subsection{Action of $GL_n(\mathbb{R})$ on $\mathrm{Sym}^{+}_n$ by congruence}
\label{congruence}

Here, we examine a suitable Lie group action to build Lie group integration schemes on $\mathrm{Sym}^{+}_n$. First, $\mathrm{Sym}^{+}_n$ is indeed a smooth manifold, whose tangent space at each point can be identified with the set of symmetric matrices $\mathrm{Sym}_{n}$. Thus, any differential equation on $\mathrm{Sym}^{+}_n$ has a symmetric matrix as a RHS. \\
We choose the Lie group to be the general linear group $GL_n(\mathbb{R})$. Its Lie algebra is simply $\mathbb{R}^{n\times n}$. The group action we consider is:
\begin{align}
\begin{split}
\boldsymbol{\Lambda}:\ GL_n(\mathbb{R}) \times \mathrm{Sym}^{+}_n &\rightarrow \ \mathrm{Sym}^{+}_n \\
 (\mathbf{M}, \mathbf{P}) &\mapsto \mathbf{MPM}^T.
\end{split}
\label{groupaction}
\end{align}
It is a well known group action, and we can easily check it is indeed transitive. For covariance matrices, it corresponds to the effect of an invertible linear transformation of a random vector on its covariance matrix.\\
Using~\eqref{homomorphism}, we can derive the algebra action $\boldsymbol{\lambda}_{*}: \mathbb{R}^{n\times n}\times \mathrm{Sym}^{+}_n  \rightarrow \ \frak{X}(\mathrm{Sym}_{n}^{+})$:
    \begin{equation}
        \boldsymbol{\lambda}_{*}(\mathbf{M})(\mathbf{P}) = \frac{d}{ds} \boldsymbol{\Lambda}(\mathbf{Y}(s),\mathbf{P}) |_{s=0} = \mathbf{M}\mathbf{P} + \mathbf{P}\mathbf{M}^{T},
        \label{lambdastar}
    \end{equation}
where $\mathbf{Y}(s)=\mathbf{I}+ s\mathbf{M} + ...$ is a smooth curve on the Lie group with $\mathbf{Y}(0) = \mathbf{I}$ and initial speed $\mathbf{M} \in \mathbb{R}^{n \times n}$. Following Eq.~\eqref{flow}, we can tackle equations of the form
    \begin{equation} 
\frac{d\mathbf{P}}{dt} = \boldsymbol{\xi}(\mathbf{P},t)\mathbf{P} +  \mathbf{P}\boldsymbol{\xi}(\mathbf{P},t)^{T},
\label{lie_cov}
\end{equation}
with $\boldsymbol{\xi} :\mathrm{Sym}^{+}_n \times \mathbb{R}^+ \rightarrow \mathbb{R}^{n \times n}$ \emph{any} smooth function. Eq.~\eqref{lie_cov} is not very restrictive and many equations of interest can be written this way.
For instance, with a constant $\boldsymbol{\xi}$, Eq.~\eqref{lie_cov} governs the dynamics of the covariance of a random variable that propagates via a deterministic linear dynamical system~\cite{pannekoucke2016parametric}. More complex functions $\boldsymbol{\xi}$ can model more complex situations.

\subsection{Action of the symplectic group on the imaginary part of the Siegel Upper Halfspace}

\subsubsection{Definitions}

Most of the basic results of this section can be found e.g. in \cite{freitas1999action}, with corresponding proofs. The Siegel Upper Halfspace is defined as the set of complex matrices

\begin{equation}
    SH_n = \{ \mathbf{X}+i\mathbf{Y}, \mathbf{X} \in \mathrm{Sym}_n, \mathbf{Y}\in \mathrm{Sym}^{+}_n\} 
\end{equation}

The Symplectic group $SP(2n)$, in the other hand, is defined as the group of matrices of size $2n \times 2n$ that preserve a certain nondegenerate skew-symmetric bilinear form (the standard symplectic form, essential in symplectic geometry that generalizes hamiltonian mechanics to smooth manifolds). The group is defined as:
\begin{equation}
    SP(2n) =  \left\{\mathbf{M} =  \begin{bmatrix}
\mathbf{A} & \mathbf{B} \\
\mathbf{C} & \mathbf{D}
\end{bmatrix} \in  \mathbb{R}^{2n\times 2n}, \mathbf{M}^T \mathbf{JM} = \mathbf{J}\ \right\}
\end{equation}
with $\mathbf{J} =  \begin{bmatrix}
\mathbf{0} & \mathbf{I}_n \\
-\mathbf{I}_n & \mathbf{0}
\end{bmatrix} \in \mathbb{R}^{2n \times 2n}$.

Writing out explicitly the condition defining $SP(2n)$, we get the following requirement:
\begin{equation}
     \begin{bmatrix}
\mathbf{A} & \mathbf{B} \\
\mathbf{C} & \mathbf{D}
\end{bmatrix} \in SP(2n) \Leftrightarrow \mathbf{A}^T\mathbf{C} \in \mathrm{Sym}_n, \mathbf{B} ^T\mathbf{D} \in \mathrm{Sym}_n \ \textrm{and} \ \mathbf{A}^T\mathbf{C} - \mathbf{B}^T \mathbf{D} = \mathbf{I}_n
\end{equation}

$SP(2n)$ is a Lie group, whose Lie algebra is the set of $2n \times 2n$ matrices $\mathfrak{sp}(2n)$ verifying
\begin{equation}
    \mathbf{M}\in \mathbb{R}^{2n}, \mathbf{M}^T\mathbf{J}+ \mathbf{JM} = \mathbf{0}.
\end{equation}
This condition translates to the block representation as
\begin{equation}
    \mathbf{M} = \begin{bmatrix}
\mathbf{A} & \mathbf{B} \\
\mathbf{C} & \mathbf{-A}^T
\end{bmatrix}, \ \textrm{with} \  \mathbf{B},\mathbf{C} \in  \mathrm{Sym}_n
\end{equation}

\subsubsection{Framework}
Now let us define 

\begin{align}
\begin{split}
\boldsymbol{\Lambda}:\ SP(2n) \times  SH_n &\rightarrow \  SH_n \\
 \left( \mathbf{M} = \begin{bmatrix}
\mathbf{A} & \mathbf{B} \\
\mathbf{C} & \mathbf{D}
\end{bmatrix} , \mathbf{Z} \right) &\mapsto (\mathbf{AZ}+\mathbf{B})(\mathbf{CZ}+\mathbf{D})^{-1}.
\end{split}
\label{groupactionsiegel}
\end{align}

This function actually defines a transitive group action of $SP(2n)$ to $SH_n$. There are a number of requirements to check: first that $\boldsymbol{\Lambda}(\mathbf{M},\mathbf{Z}) \in SH_n$, that this operation actually defines a group action, and finally transitivity. All proofs can be found in~\cite{freitas1999action}.\\

Now we can derive the corresponding Lie algebra Homomorphism, which, following~\eqref{homomorphism}, writes (proof omitted here):

\begin{equation}
        \boldsymbol{\lambda}_{*}( \mathbf{M})(\mathbf{Z}) = \frac{d}{ds} \Lambda(\boldsymbol{\rho}(s),\mathbf{Z}) |_{s=0} = \mathbf{AZ}+\mathbf{ZA}^T  + \mathbf{B} - \mathbf{ZCZ} .
\end{equation}

with $\boldsymbol{\rho}(0) = \mathbf{I}$ and $\boldsymbol{\rho}'(0) = \mathbf{M}$.This means that with this homogeneous space representation of $\mathrm{Sym}^{+}_n$ (only considering imaginary parts), we can tackle ODEs of the form

\begin{equation}
    \frac{d\mathbf{P}}{dt} = \mathbf{A}(\mathbf{P},t)\mathbf{P} + \mathbf{PA}(\mathbf{P},t)^T + \mathbf{B}(\mathbf{P},t) - \mathbf{PC}(\mathbf{P},t)\mathbf{P}
\end{equation}
Where $\mathbf{A}(\mathbf{P},t)$, $\mathbf{B}(\mathbf{P},t)$, $\mathbf{C}(\mathbf{P},t)$ are smooth functions of $\mathbf{P}\in \mathrm{Sym}^{+}_n$ and time, with values in $\mathbb{R}^{n \times n}$, $ \mathrm{Sym}_n$ and $ \mathrm{Sym}_n$, respectively. \\

This result is remarkable because the ODEs we can tackle are matrix and time dependent versions of the Riccati equation in LQ control, or to evolve the covariance in a Kalman-Bucy filter. They also encompass the covariance equation for the multivariate geometric Brownian motion (see next section for details).\\

Even though theoretically, this group action is more powerful than that of section~\ref{congruence} since it allows to tackle more general equations, mumerically speaking, the action of $GL_n(\mathbb{R})$ by congruence is more stable (most likely because simply computing the action of symplectic matrices already requires a matrix inversion). Hence we only consider the latter in the experiments of the next section. 
\section{Case Study}
\label{sec:exp}
\vspace{-0.15cm}
\subsection{Multivariate Geometric Brownian Motion}
\begin{figure}
\vspace{-0.4cm}
    \centering
          \includegraphics[scale = 0.22]{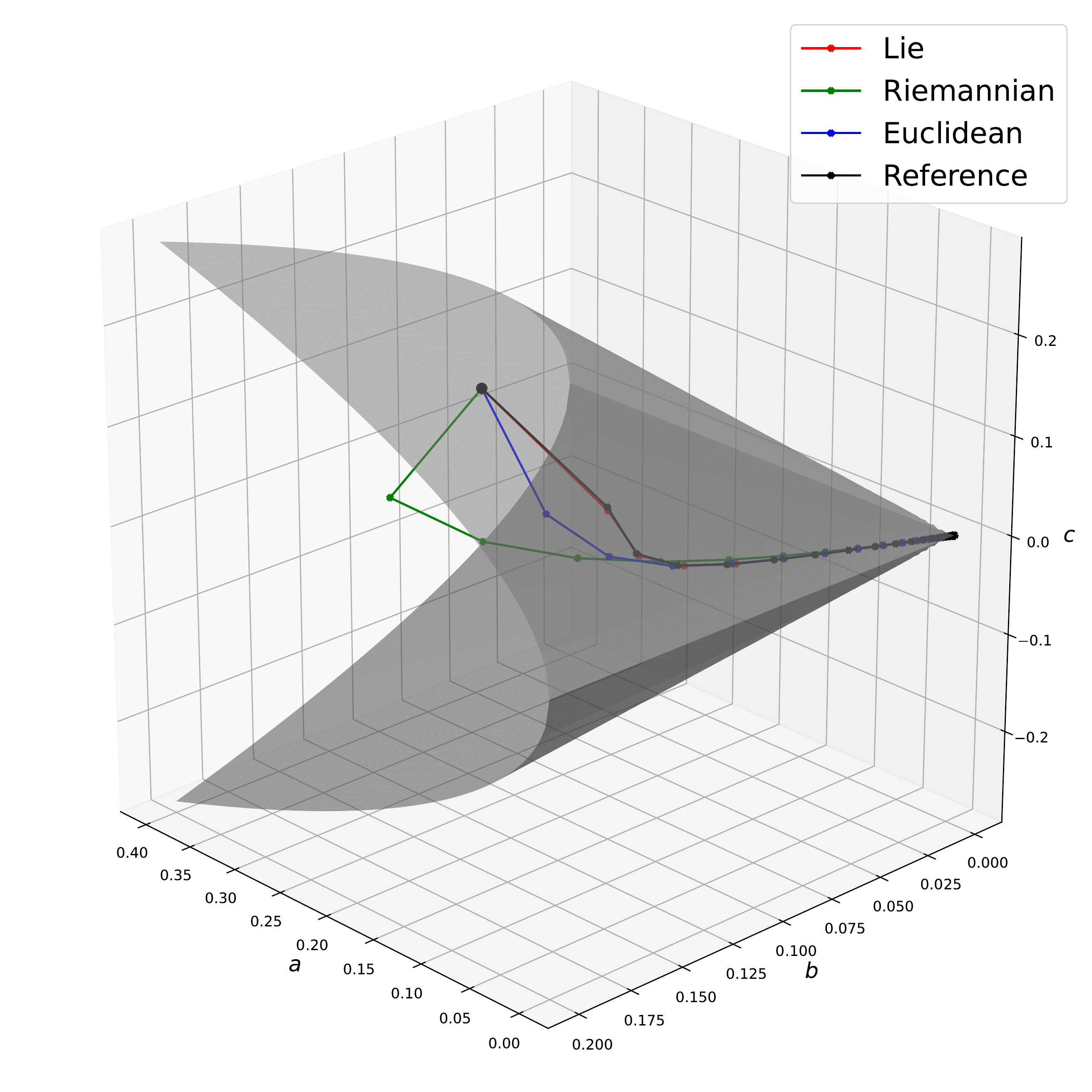}   
               \includegraphics[scale = 0.22]{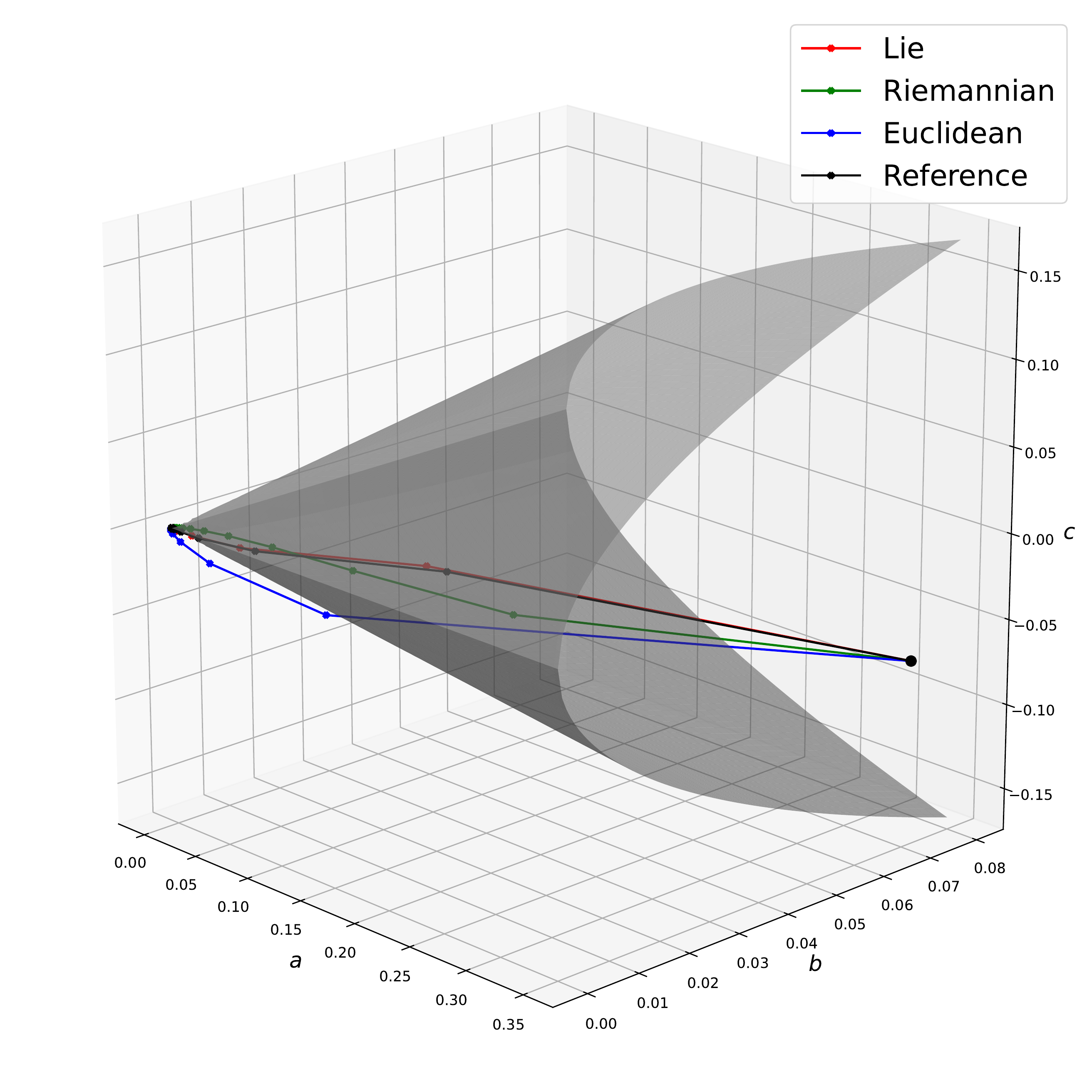}   
         \caption{Trajectories of the covariance matrices in $\mathrm{Sym}^{+}_2$ obtained by the different RK4 methods for Eq.~\eqref{cov_gbm} using the embedding
$\left(\protect\begin{smallmatrix}a&c\\c&b\protect\end{smallmatrix}\right) \mapsto (a,b,c)\in \mathbb{R}^3$. The boundary of $\mathrm{Sym}^{+}_2$ is given by the surface (in gray) defined by $ab-c^2 = 0$. Left: first case, where the iterates always stay in the manifold but only the Lie-RK4 method is accurate. Right: second case, where the Euclidean method produces iterates outside  $\mathrm{Sym}^{+}_2$.}
    \label{fig:traj}
    \vspace{-0.4cm}
\end{figure}
We are interested here in a multivariate generalization of a Geometric Brownian Motion (GBM), given by the (Itô) SDE~\cite{barrera2022cutoff}:
\begin{equation}
    d\rvx{X} = \left( \mathbf{A}+\frac{1}{2}  \mathbf{B}^2 \right) \rvx{X} dt + \mathbf{B}\rvx{X} dW_t,
    \label{SDE}
\end{equation}
with $\rvx{X}$ a random vector of size $n$, $\mathbf{A},\mathbf{B} \in \mathbb{R}^{n\times n}$ two commuting matrices, such that the eigenvalues of $\mathbf{A}+\frac{1}{2} \mathbf{B}^2$ have a strictly negative real part. $W_t \in \mathbb{R}$ is a Brownian motion. A closed form expression of the trajectories for a deterministic initial condition $\mathbf{x}_0$ is given by $\rvx{X}(t) = \exp(t\mathbf{A} + \mathbf{B}W_t)\mathbf{x}_0$.
We can derive ODEs followed by the mean $\mathbf{m}$ (taking expectations in~\eqref{SDE}) and covariance matrix $\mathbf{P}$ (using Itô's Lemma~\cite{oksendal2013stochastic} on $\rvx{X}\rvx{X}^T$, taking expectations, and a few algebraic manipulations, see~\cite{arxiv_version}). They provide a broad summary of the statistics of the process up to order two (much simpler than e.g. the Fokker-Planck equation):
\begin{align}
    &\frac{d\mathbf{m}}{dt} = \left( \mathbf{A}+\frac{1}{2}  \mathbf{B}^2 \right) \mathbf{m} \\
    &\frac{d\mathbf{P}}{dt} = \left( \mathbf{A}+\frac{1}{2}  \mathbf{B}^2 \right) \mathbf{P} + \mathbf{P}\left( \mathbf{A}+\frac{1}{2}  \mathbf{B}^2 \right)^{T}  \! \! +\mathbf{B}(\mathbf{P} + \mathbf{m}\mathbf{m}^T)\mathbf{B}^{T}.
 \label{cov_gbm}
\end{align}
Eq.~\eqref{cov_gbm} can indeed be put in the form of Eq.~\eqref{lie_cov}, using 
\begin{equation}
\boldsymbol{\xi}(\mathbf{P}) = \left( \mathbf{A} + \frac{1}{2} \mathbf{B}^2 \right) + \frac{1}{2} \mathbf{B} ( \mathbf{P} +\mathbf{m}\mathbf{m}^{T})\mathbf{B}^T\mathbf{P}^{-1}.
\label{xigbm}
\end{equation}
We can only detail this example (with $n=2$) here, but our method also applies e.g. to the covariance of a multivariate Ornstein-Uhlenbeck process~\cite{meucci2009review}, or to various Riccati equations encountered in control~\cite{bacsar1998dynamic}, with suitable choices of $\boldsymbol{\xi}$. Still, we will consider two cases, only differing in the way the matrix $\mathbf{A}$ is defined, and the chosen time step, all other parameters remaining the same.
\subsection{Choice of numerical values for the SDE~\eqref{SDE}}
For the case study of Section~\ref{sec:exp}, we chose $n=2$ to be able to visualize the trajectories in the natural embedding of $\mathrm{Sym}_{2}^{+}$ in $\mathbb{R}^3$, see Fig~\ref{fig:traj}. $\mathbf{P}_0$ was generated by using $\mathtt{scikit-learn}$'s function $\mathtt{make\textunderscore spd\textunderscore matrix}$~\cite{pedregosa2011scikit}, with the same random state in both cases, and multiplying the result by a factor of 0.15. In practice, we have $\mathbf{P}_0 \approx \begin{pmatrix}
0.3383 & -0.0716  \\
-0.0716 & 0.0743  
\end{pmatrix}$. We chose commuting $\mathbf{A}$ and $\mathbf{B}$ matrices so as to be able to use the closed form solution given in~\cite{barrera2022cutoff}. We further chose them such that $\mathbf{A} + \frac{1}{2}\mathbf{B}^2$ is negative definite. More in detail, we chose a symmetric $\mathbf{B} = \begin{pmatrix}
-0.4 & 0.1  \\
0.1 & -0.2  
\end{pmatrix}$
To make sure  $\mathbf{A}$ and $\mathbf{B}$ commute, we chose  $\mathbf{A}$ by diagonalizing $\mathbf{B} = \mathbf{ODO}^T$, changing the eigenvalues in $\mathbf{D}$, and then computing $\mathbf{A} = \mathbf{OD'O}^T$. In the first case, we chose $\mathbf{D'} = \begin{pmatrix}
-10 & 0  \\
0 & -2  
\end{pmatrix}$, and in the second case we chose $\mathbf{D'} = \begin{pmatrix}
-4 & 0  \\
0 & -8  
\end{pmatrix}$.\\
We set $t \in [0,2]$ (resp. $t \in [0,1.5]$), and show results with $30$ (resp. $11)$ evenly spaced time steps in the first (resp. second) case. In the second case, we note that this yields a time step $h = 0.15$. Using the eigenvalues of $\mathbf{P}_0$ and of the RHS of eq.~\eqref{cov_gbm}, we obtain $\rho_{\textrm{max}} = 0.02121$, and $\rho_\textrm{min} = 0.1345$. For the RK4 algorithm, similarly, we obtain $\rho_\textrm{max} = 0.0223$ and $\rho_\textrm{min} = 0.1414$. With our choice of time step, the classical Euler and RK4 methods will then fail in the second case. In practice, the upper bound $\rho_\textrm{max}$ seems quite loose on that example and it seems experimentally that the iterate stays on the manifold as long as $\rho \leq \rho_\textrm{min}$.
\vspace{-0.3cm}
\subsection{Results}
\begin{figure}
    \centering
        \includegraphics[scale = 0.19]{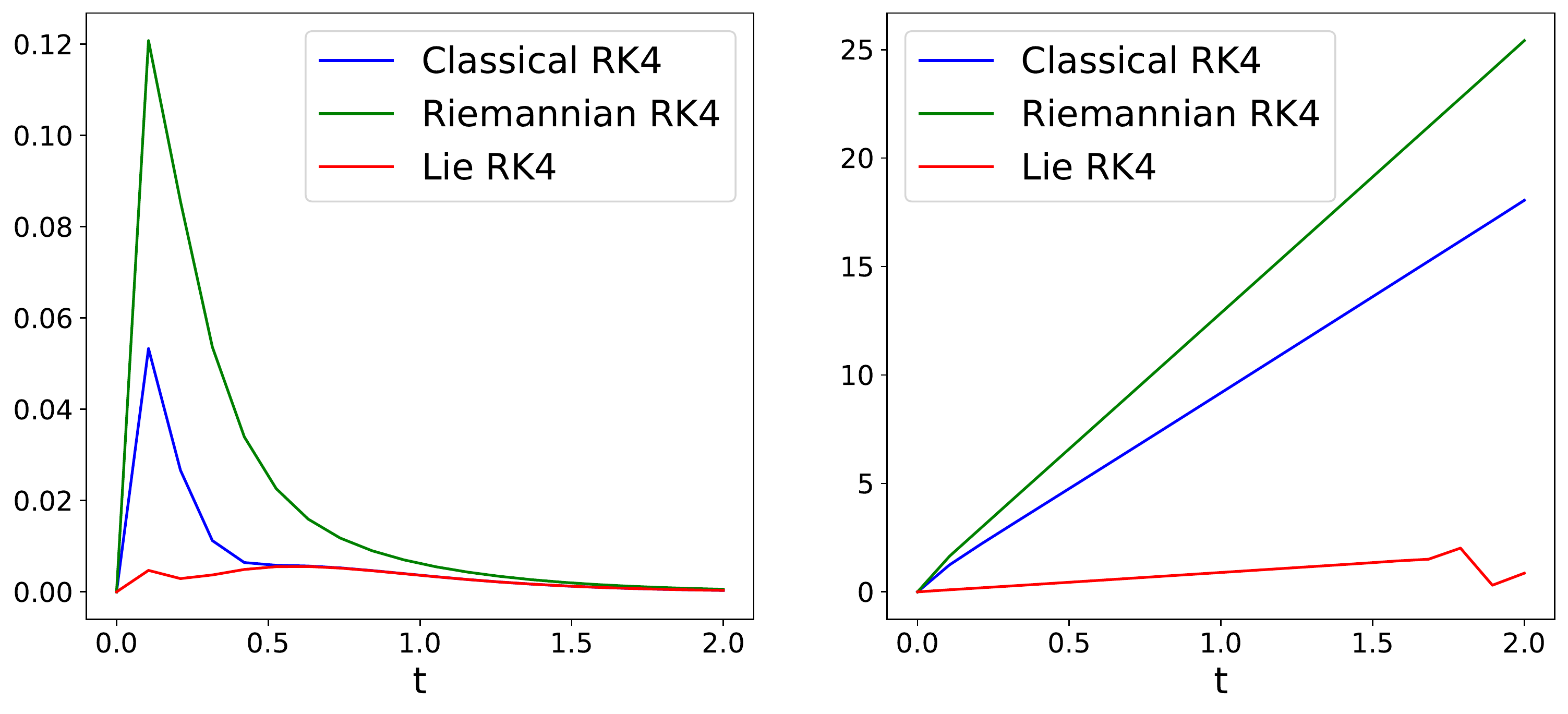}\rulesep
     \includegraphics[scale = 0.19]{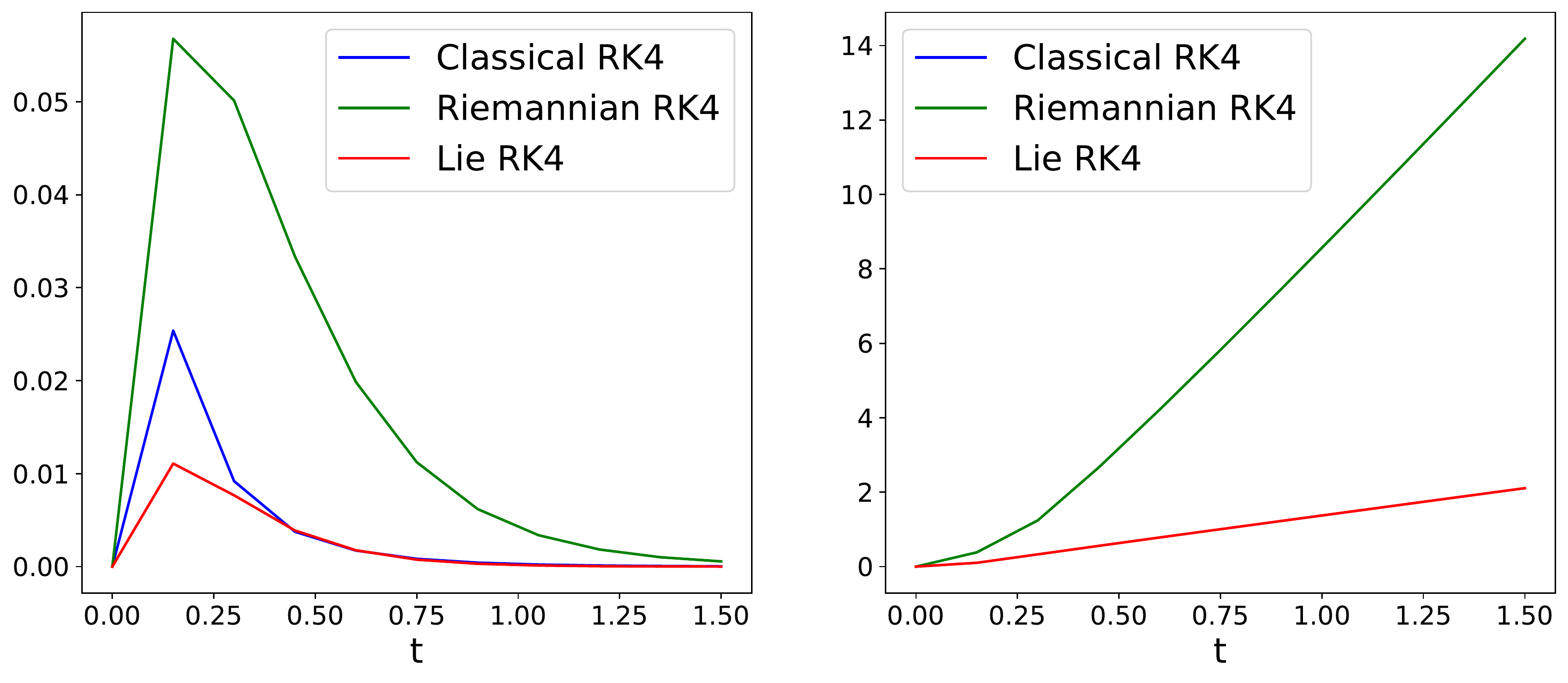}
    \caption{Euclidean and Riemannian distances between each integrated trajectory and the reference, on the left and right of each panel, respectively. The case corresponding to the left of Fig.~\ref{fig:traj} is on the left panel, and the other case is in the right panel. In the latter, the Riemannian distance cannot be computed for the Euclidean method since the iterates are outside the manifold.}
    \label{fig:dist}
\end{figure}

We compare three explicit RK4 schemes for Eq.~\eqref{cov_gbm}: a Euclidean scheme, a variant where each step is brought back to the manifold using a Riemannian exponential map, with the affine invariant metric of~\cite{bhatia2009positive} (Riemmanian RK4), and our method (Lie RK4). We compare them to a reference obtained from a classical RK4 method with a small enough time step so we know the trajectory remains on $\mathrm{Sym}_{n}^{+}$ and the integration error is small. We start from a Gaussian initial condition $\rvx{X}_0 \sim \mathcal{N}(\mathbf{m}_0,\mathbf{P}_{0})$. Then, the process converges to a distribution given by a Dirac centered at $\mathbf{0}$. We show the trajectories obtained for the three competing methods embedded in $\mathbb{R}^3$ in Fig.~\ref{fig:traj} (left: first case, right: second case), and plot two distances between the trajectories and the reference in Fig.~\ref{fig:dist}: the Euclidean distance in $\mathbb{R}^{n \times n}$, as well as the affine invariant distance~\cite{bhatia2009positive} on $\mathrm{Sym}^{+}_2$.\\
We see from both Figs~\ref{fig:traj} and~\ref{fig:dist} that in the first case, only the Lie-RK4 method is able to accurately follow the reference dynamics for this value of the integration step. The Euclidean RK4 method is unable to produce a good solution while still producing iterates remaining on the manifold, which may be quite deceptive in applications. The Riemannian RK4 method here actually performs worse than its Euclidean counterpart. Indeed, thus method is not truly intrinsic to the manifold, so a bad step in the Euclidean domain often cannot be completely made up for. In the second case, the configuration is different and the integration step larger, so the result is that Euclidean RK4 produces iterates that immediately leave the manifold, producing a statistically meaningless trajectory. This is why the Riemannian distance cannot be computed for this method in the right panel of Fig~\ref{fig:dist}. In this case, Riemannian RK4 is able to guarantee that the iterates remain in the manifold, but the solution does not follow accurately the true dynamics. Finally, Lie-RK4 produces a very accurate solution guaranteed to stay in the manifold, even in such a challenging situation.\\
For smaller time steps, Lie RK4 remains the most precise up to a point, then classical RK4 may becomes marginally better than our approach in some cases, probably due to accumulating errors during the additional computations. However, in many applications, the step size is imposed by the problem, e.g. when observation data have a low sampling rate.
\vspace{-0.2cm}
\section{Conclusion}
\label{sec:ccl}
We have presented a Lie group framework to define structure-preserving integration schemes for flows of SPD matrices. Our fully intrinsic integrators keep iterates on the manifold, and provides smaller integration error than classical or naive methods, especially for large time steps. This will be useful in our future work to learn and represent uncertainty in data assimilation~\cite{dridi2021learning,nguyen2020variational} or controls from observation data when governing equations are unknown (by learning a function $\boldsymbol{\xi}$ matching the data). In such cases, the time step is imposed by data and the training process may lead to ill-conditioned equations.

\section{Appendices}

\subsection{Proof of Theorem~\ref{th}}

In this section, we prove Theorem~\ref{th}, which provides a sufficient condition on the integration step $\rho$ for the classical explicit Euler method (or any explicit Runge-Kutta method, for that matter) to yield an iterate that does not belong to $\mathrm{Sym}_n^+$, in the case where the (necessarily symmetric) RHS of the ODE evaluated at the current point has at least one negative eigenvalue. $\mathbf{T}$ has to be understood to be the right hand side of the ODE, computed for the current value of the matrix to be integrated. Besides, we also provide a condition on the time step to guarantee that the next iterate of a classical Euler method remains on the manifold. If explicit RK methods are used instead of explicit Euler methods, then $\mathbf{T}$ denotes the (symmetric) vector field used in the RK update instead and the results still apply. 

\begin{proof}

\begin{enumerate}
    \item if $\mathbf{T}$ is positive semidefinite, $\forall \mathbf{x} \in \mathbb{R}^{n}$ we have
    \begin{equation*}
        \mathbf{x}^{T}(\mathbf{P} + \rho \mathbf{T}) \mathbf{x} = \mathbf{x}^T \mathbf{Px} + \rho \mathbf{x}^T \mathbf{Tx} > 0,
    \end{equation*}
    since $\mathbf{x}^T \mathbf{Tx}\geq 0$. Then $\mathbf{P} + \rho \mathbf{T}\in \mathrm{Sym}_{n}^{+}$ for any $\rho \geq 0$.
    \item if $\mathbf{T}$ has at least one negative eigenvalue, we use inequalities due to Weyl (1912), \cite{weyl1912asymptotische,tao} bounding the eigenvalues of a sum of symmetric matrices by sums of suitable pairs of eigenvalues of each matrix:
    \begin{lemma}[Weyl inequalities~\cite{weyl1912asymptotische}]
    Let $\mathbf{N},\mathbf{R}\in \mathrm{Sym}_{n}$, and $\mathbf{M} = \mathbf{R}+\mathbf{N}$. Then denoting as $\lambda_1 \leq ... \leq \lambda_n$ the eigenvalues of $\mathbf{R}$, $\nu_1 \leq ... \leq \nu_n $ the eigenvalues of $\mathbf{N}$, and as $\eta_1 \leq ... \leq \eta_n$ the eigenvalues of $\mathbf{M}$, we have 
    \begin{enumerate}
        \item For any $1\leq i,j,k \leq n$ such that $i+j = n+k$:
        \begin{equation}
        \eta_k \leq \lambda_i + \nu_j.
        \label{firstweyl}
        \end{equation}
        \item For any $1\leq i,j,k \leq n$ such that $i+j = 1+k$:
        \begin{equation}
        \lambda_i + \nu_j \leq \eta_k.
        \label{secondweyl}
        \end{equation}
        
    \end{enumerate}
    \end{lemma}
In our case, we are interested in knowing whether the smallest eigenvalue of $\mathbf{P} + \rho \mathbf{T}$ is negative or not (i.e. when at least one eigenvalue is negative). Thus, we apply the Weyl inequalities~\eqref{firstweyl} with $\mathbf{R} = \mathbf{P}$ and $\mathbf{N} = \rho \mathbf{T}$, with $k = 1$, denoting here as $\nu_1 \leq ... \leq \nu_n $ the eigenvalues of $\mathbf{T}$, and we obtain
\begin{equation*}
   \eta_1  \leq \lambda_i + \rho \nu_j.
\end{equation*}
for any $(i,j)$ such that $i+j \leq n+1$. 
If the RHS is smaller than 0, then $\mathbf{P} + \rho \mathbf{T} \notin \mathrm{Sym}_{n}^{+}$. Thus, the only case for which the bound is interesting are those for which $\nu_j \leq 0$, otherwise the RHS is positive. From this, requiring the RHS to be negative yields the following lower bound on $\rho$:
    \begin{equation*}
    \rho \geq -\frac{\lambda_j}{\nu_i},
\end{equation*}
In particular, we are interested in the the tightest bound, so in the tuple $(i,j) \in S$ resulting in the smallest RHS, hence the minimum in the statement of the theorem.
\item We prove this point by contraposition. Let us assume that $\mathbf{P}+\rho\mathbf{T} \notin \mathrm{Sym}_{n}^{+}$. Then $\eta_1 \leq 0$ (at least). This time, we use the second set of Weyl inequalities~\eqref{secondweyl} with $k = 1$. Necessarily, we need $i=j=1$ (the smallest eigenvalues of $\mathbf{P}$ and $\mathbf{T}$) so that $i+j = 1+k=2$. We obtain
    \begin{equation*}
        \lambda_1 + \rho \nu_1 \leq \eta_1 \leq 0.
    \end{equation*}
    Again, isolating $\rho$ yields:
    $\rho \geq -\frac{\lambda_1}{\nu_1}$. 
\end{enumerate}
\end{proof}


\subsection{Proof of the transitivity of the group action~\eqref{groupaction}}

\begin{proof}
Let $\mathbf{X},\mathbf{Y} \in \textrm{Sym}_{n}^{+}$, and define $\mathbf{G} = \mathbf{Y}^{\frac{1}{2}}\mathbf{X}^{\frac{-1}{2}} \in {GL}_n (\mathbb{R})$. Then
\begin{align*}
    \boldsymbol{\Lambda}(\mathbf{G},\mathbf{X}) = \mathbf{GXG}^T  & = \mathbf{Y}^{\frac{1}{2}}\mathbf{X}^{\frac{-1}{2}} \mathbf{X} (\mathbf{Y}^{\frac{1}{2}}\mathbf{X}^{\frac{-1}{2}})^T \\
    &=  \mathbf{Y}^{\frac{1}{2}}\mathbf{X}^{\frac{-1}{2}} \mathbf{X}^{\frac{1}{2}}\mathbf{X}^{\frac{1}{2}} \mathbf{X}^{\frac{-1}{2}}\mathbf{Y}^{\frac{1}{2}}\\
    & = \mathbf{Y}^{\frac{1}{2}} \mathbf{I} \mathbf{Y}^{\frac{1}{2}} \\
    & = \mathbf{Y}.
\end{align*}
\end{proof}
\subsection{Computation of the Lie algebra homomorphism~\eqref{lambdastar}}

We start from the definition given in Eq.~\eqref{homomorphism}, and use the fact that on a matrix group such as $GL_n(\mathbb{R})$, a curve $\boldsymbol{\rho}$ for which $\boldsymbol{\rho}(0) = \mathbf{I}$ has an initial speed that can be defined in the usual Euclidean sense, i.e. $\boldsymbol{\rho}'(0) = \mathbf{M}$, with $\mathbf{M}\in \mathcal{M}_{n}(\mathbb{R})$. Then, the curve can be written as a Taylor expansion at zero:
\begin{equation*}
    \boldsymbol{\rho}(s)=\mathbf{I}+ s\mathbf{M} + o(s).
\end{equation*}

Then, working from the definition we have
\begin{align*}
    \boldsymbol{\lambda}_{*}(\mathbf{M})(\mathbf{P}) & \triangleq \frac{d}{ds} \boldsymbol{\Lambda}(\boldsymbol{\rho}(s),\mathbf{P}) |_{s=0} \\
    & = \frac{d}{ds} (\boldsymbol{\rho}(s)\mathbf{P} \boldsymbol{\rho}(s)^T)|_{s=0} \\
    & = \frac{d}{ds} ((\mathbf{I}+ s\mathbf{M} + o(s))\mathbf{P} (\mathbf{I}+ s\mathbf{M} + o(s))^T)|_{s=0} \\
    & = (\mathbf{M}+o(1)) \mathbf{P}  (\mathbf{I}+ s\mathbf{M} + o(s))   +  (\mathbf{I}+ s\mathbf{M} + o(s))^T \mathbf{P}(\mathbf{M}+o(1))^{T}|_{s=0}\\
    & = \mathbf{MP} + \mathbf{PM}^T.
\end{align*}

\subsection{Derivation of the covariance equations and corresponding $\boldsymbol{\xi}$ for several equations of interest} 

Here we derive, for several examples of interest, equations followed by covariance matrices. The first three cases concern SDE on a stochastic process $\rvx{X}_t\in \mathbb{R}^{n}$ for which we are interested in the evolution of the covariance matrix  (we drop the time index for brevity)\begin{equation*}
    \mathbf{P} = \mathbb{E}\left[(\rvx{X}-\mathbb{E}[\rvx{X}])(\rvx{X}-\mathbb{E}[\rvx{X}])^T \right] = \mathbb{E}[\rvx{XX}^T] - \mathbb{E}[\rvx{X}]\mathbb{E}[\rvx{X}]^T.
\end{equation*}
The last case is different in nature and concerns an optimal control problem in continuous time whose solution involves solving an equation on an SPD matrix.

\subsubsection{Linear deterministic system with stochastic initial condition}

The first equation of interest is simply a linear deterministic dynamical system, given by
\begin{equation*}
    \frac{d\rvx{X}}{dt} = \mathbf{A} \rvx{X}.
\end{equation*}
This describes a stochastic process if the initial condition is given by a probability distribution instead of a deterministic value. In that case, taking the expectation of the solutions for any possible $\rvx{X}_{0}$, we get the same ODE on $\mathbf{m}(t) = \mathbb{E}[\rvx{X}(t)]$.
\begin{equation*}
        \frac{d\mathbf{m}}{dt} = \mathbf{A} \mathbf{m}.
\end{equation*}
Then, to obtain a differential equation on $\mathbf{P}$, we first compute the derivative of $\rvx{X}\rvx{X}^T$:
\begin{equation*}
    \frac{d(\rvx{XX}^T)}{dt} = \frac{d\rvx{X}}{dt}\rvx{X}^T + \rvx{X}\frac{d\rvx{X}^T}{dt} = \mathbf{A}\rvx{XX}^T + \rvx{XX}^T\mathbf{A}^T.
\end{equation*}
Then we can obtain, taking expectations in the previous equation, subtracting $ \mathbf{mm}^T $ to form the covariance (we can swap derivation and expectation by the Leibniz integral rule for Itô diffusions, since $\rvx{X}_t$ is absolutely continuous for any $t>0$ and its density is assumed to be smooth enough in both $\mathbf{x}$ and $t$):
\begin{align*}
    \frac{d\mathbf{P}}{dt} & = \frac{d}{dt} \left( \mathbb{E}[ \rvx{X}\rvx{X}^T ] - \mathbf{mm}^T \right)  \\
    & = \mathbf{A} \mathbb{E}[\rvx{XX}^T] +\mathbb{E}[\rvx{XX}^T]\mathbf{A}^T - \mathbf{A}\mathbf{mm}^T - \mathbf{mm}^T \mathbf{A}^T \\
    & = \mathbf{A}  \left( \mathbb{E}[ \rvx{X}\rvx{X}^T ] - \mathbf{mm}^T \right)  +  \left( \mathbb{E}[ \rvx{X}\rvx{X}^T ] - \mathbf{mm}^T \right) \mathbf{A}^T \\
    & = \mathbf{AP} + \mathbf{PA}^T.
\end{align*}
From this, it is clear that taking $\boldsymbol{\xi}(\mathbf{P},t) \equiv \mathbf{A}$ in~\eqref{lie_cov} corresponds exactly to this equation. In other words, the simplest ODE the group action~\eqref{groupaction} can handle models the evolution of the covariance matrix of a process passing through a linear dynamical system.
\subsubsection{Multivariate Ornstein-Uhlenbeck process}

We now switch to an actual (Itô) SDE, in this case a linear one with a constant diffusion:
\begin{equation*}
d \rvx{X}_t =  \mathbf{A}\rvx{X}_tdt + \mathbf{B} d\rvx{W}_t,
\end{equation*}
where $\rvx{X}\in \mathbb{R}^{n}$, $ \mathbf{A} \in \mathbb{R}^{n\times n}$, $\mathbf{B}  \in \mathbb{R}^{n\times n}$, $\rvx{W}_t \in \mathbb{R}^{n}$ is a multivariate Brownian motion (with independent entries). This is a multivariate generalization of the well-known Ornstein-Uhlenbeck process. Since the expectation of the Brownian motion is zero, the ODE followed by the mean of the process is the same as in the previous example. This entails that the mean of the process will converge to zero as well as long as $\mathbf{A}$ has no eigenvalues with a positive real part. There is a closed form solution for both the trajectories and the covariance matrix for a deterministic initial condition (see. e.g.~\cite{meucci2009review}). Even if the initial condition is stochastic, we can derive the ODE governing the evolution of the covariance of the process. To do this, we must first derive the SDE followed by $\rvx{XX}^T$, using Itô's Lemma~\cite{oksendal2013stochastic}. We have $(\rvx{XX}^T)_{ij} = X_i X_j$. Applying Itô's Lemma to this function yields:
\begin{align*}
    d(X_i X_j) & = \sum_{k=1}^{n} \frac{\partial (X_i X_j)}{\partial  X_k} dX_k + \frac{1}{2} \sum_{k=1}^{n} \sum_{l=1}^{n} \frac{\partial^2 (X_i X_j)}{\partial X_k \partial X_l} dX_k dX_l \\
    & = X_j dX_i + X_i dX_j + dX_i dX_j.
\end{align*}
Hence, gathering all these terms for $1\leq i,j \leq n$ in a matrix form, we get:
\begin{equation*}
    d(\rvx{X}\rvx{X}^T) =(d \rvx{X}) \rvx{X}^T  + \rvx{X}d\rvx{X}^T + d\rvx{X}d\rvx{X}^T.
\end{equation*}
Replacing $d \rvx{X}$ with its expression, and expanding, we obtain:
\begin{align*}
     d(\rvx{X}\rvx{X}^T) & = (\mathbf{A}\rvx{X}dt + \mathbf{B} d\rvx{W}_t)\rvx{X}^T + \rvx{X}(\mathbf{A}\rvx{X}dt + \mathbf{B} d\rvx{W}_t)^T  \\
     &  \quad \quad \quad \quad + (\mathbf{A}\rvx{X}dt + \mathbf{B} d\rvx{W}_t)(\mathbf{A}\rvx{X}dt + \mathbf{B} d\rvx{W}_t)^T \\
     & = \mathbf{A} \rvx{XX}^T dt + \mathbf{B}d\rvx{W}_t \rvx{X}^T + \rvx{XX}^T \mathbf{A}^T dt \\
     & + \rvx{X} d\rvx{W}_t^T \mathbf{B}^T + \mathbf{A} \rvx{XX}^T\mathbf{A}^T dt^2  + \mathbf{A} \rvx{X}dt d\rvx{W}_t^T \mathbf{B}^T \\
     & + \mathbf{B}d\rvx{W}_t \mathbf{A} \rvx{X}dt   + \mathbf{B}d\rvx{W}_t d\rvx{W}_t \mathbf{B}^T.
\end{align*}
Using the usual multiplication rules, $dt^2 = dt d\rvx{W}_t = 0 $ and $d\rvx{W}_t  d\rvx{W}_t^T = dt \mathbf{I}$, we obtain
\begin{align*}
    d(\rvx{XX}^T) & = (\mathbf{A}\rvx{XX}^T + \rvx{XX}^T \mathbf{A}^T + \mathbf{BB}^T) dt +\rvx{X} d\rvx{W}_t^T \mathbf{B}^T +\mathbf{B}d\rvx{W}_t \rvx{X}^T.
\end{align*}
Note that this is still an Itô SDE, since by vectorizing $\rvx{XX}^T$ into a $n^2$ dimensional vector, the last SDE can be rewrittten in the usual form. To do this, we use the well known property of the Kronecker product (denoted as $\otimes$): $\mathrm{vec}(\mathbf{BVA^T}) = (\mathbf{A}\otimes \mathbf{B}) \mathrm{vec}(\mathbf{V})$, with $\textrm{vec}$ the vectorization operator. This yields
\begin{align*}
    d(\mathrm{vec}(\rvx{XX}^T))&  =  + (\mathbf{I} + \mathbf{K})  ((\rvx{X}\otimes \mathbf{A} + \textrm{vec}(\mathbf{BB}^T) )dt (\rvx{X}\otimes \mathbf{B}) d\rvx{W}_t),
\end{align*}
where $\mathbf{K}\in \mathbb{R}^{n^2 \times n^2}$ is the commutator matrix, i.e. the matrix such that $\mathrm{vec}(\mathbf{M}^T) = \mathbf{K}\mathrm{vec}(\mathbf{M}) $. Writing out the solution to this SDE, and taking expectations, we get:
\begin{equation*}
    \frac{d \mathbb{E}[\rvx{XX}^T]}{dt} = \mathbf{A}\mathbb{E}[\rvx{XX}^T] + \mathbb{E}[\rvx{XX}^T] \mathbf{A}^T + \mathbf{BB}^T.
\end{equation*}
Similarly as in the previous section, forming the covariance matrix yields:
\begin{align*}
    \frac{d\mathbf{P}}{dt} & = \frac{d}{dt} \left( \mathbb{E}[ \rvx{X}\rvx{X}^T ] - \mathbf{mm}^T \right)  \\
    & = \mathbf{A} \mathbb{E}[\rvx{XX}^T] +\mathbb{E}[\rvx{XX}^T]\mathbf{A}^T + \mathbf{BB}^T - \mathbf{A}\mathbf{mm}^T - \mathbf{mm}^T \mathbf{A}^T \\
    & = \mathbf{A}  \left( \mathbb{E}[ \rvx{X}\rvx{X}^T ] - \mathbf{mm}^T \right)  +  \left( \mathbb{E}[ \rvx{X}\rvx{X}^T ] - \mathbf{mm}^T \right) \mathbf{A}^T   + \mathbf{BB}^T\\
    & = \mathbf{AP} + \mathbf{PA}^T  + \mathbf{BB}^T.
\end{align*}


Finally, by defining $\boldsymbol{\xi}(\mathbf{P},t) = \mathbf{A} + \frac{1}{2} \mathbf{B} \mathbf{B}^T\mathbf{P}^{-1} $ in~\eqref{lie_cov} we can use our framework to integrate this equation.

\subsubsection{Multivariate Geometric Brownian Motion}

The multivariate GBM SDE is given by Eq.~\eqref{SDE}. When $\mathbf{A}$ and $\mathbf{B}$ commute, and $\mathbf{A} + \frac{1}{2}\mathbf{B}^2$ has no eigenvalue with a negative real part,~\cite{barrera2022cutoff} provides a closed form solution for an initial value of $\mathbf{x}_0$ under the form:
\begin{equation*}
    \rvx{X}(t) = \exp(t\mathbf{A} + \mathbf{B}W_t)\mathbf{x}_0.
\end{equation*}
With our choice of parameters (normal commuting matrices for $\mathbf{A}$ and $\mathbf{B}$, see Sec.~\ref{sec:exp}), we can obtain~\cite{barrera2022cutoff} a closed form solution on the variance of the process at each time step, and for any initial (deterministic) $\mathbf{x}_{0}$:
$\mathbb{E}[||\rvx{X}(t)||^2]$:
\begin{equation*}
 \mathbb{E}[||\rvx{X}(t)||^2] = \mathbb{E}[\rvx{X}(t)^T\rvx{X}(t)]  =  ||\exp(t\mathbf{Q})\mathbf{x}_{0}||^2,
\end{equation*} where $\mathbf{Q} = \frac{\mathbf{A}+(\mathbf{B}+\mathbf{B}^T)^2}{4}{}$. Then, regardless of the initial distribution, the process converges to $\mathbf{0}$ (with zero covariance).

Letting $\boldsymbol{\theta} \triangleq \mathbf{A}+\frac{1}{2} \mathbf{B}^2$, the ODE followed by $\mathbf{m}(t)$ is
\begin{equation*}
\frac{d\mathbf{m}}{dt} = \boldsymbol{\theta} \mathbf{m}.
\end{equation*}
To obtain the equation on the covariance matrix, we follow the same method as in the previous section, and from the expression of $d(\rvx{X}\rvx{X}^T)$ given by Itô's Lemma, we can compute:
\begin{align*}
     d(\rvx{X}\rvx{X}^T) & = (\boldsymbol{\theta}\rvx{X}dt + \mathbf{B} \rvx{X} dW_t)\rvx{X}^T + \rvx{X}(\boldsymbol{\theta}\rvx{X}dt + \mathbf{B}\rvx{X} dW_t)^T  \\
     &  \quad \quad \quad \quad +(\boldsymbol{\theta}\rvx{X}dt + \mathbf{B}\rvx{X} dW_t)(\boldsymbol{\theta}\rvx{X}dt + \mathbf{B}\rvx{X} dW_t)^T \\
     & =  \boldsymbol{\theta} \rvx{XX}^T dt + \mathbf{B}\rvx{X}dW_t \rvx{X}^T + \rvx{XX}^T\boldsymbol{\theta}^T dt  \\
     & + \! \rvx{X} dW_t \rvx{X}^T \mathbf{B}^T \! + \boldsymbol{\theta} \rvx{XX}^T\boldsymbol{\theta}^T dt^2 \! +\boldsymbol{\theta} \rvx{X}dt dW_t \rvx{X}^T\mathbf{B}^T \\
     & + \mathbf{B}\rvx{X} dW_t dt  \rvx{X}^T \boldsymbol{\theta}^T  + \mathbf{B} \rvx{X} dW_t^2 \rvx{X}^T \mathbf{B}^T,
\end{align*}
which yields, after applying the conventions mentioned above:
\begin{align*}
    d(\rvx{XX}^T) & = (\boldsymbol{\theta}\rvx{XX}^T + \rvx{XX}^T \boldsymbol{\theta}^T + \mathbf{B}\rvx{XX}^T\mathbf{B}^T) dt + (\mathbf{B}\rvx{XX}^T + \rvx{XX}^T\mathbf{B}^T) dW_t.
\end{align*}

Writing out the solution and taking the expectation:
\begin{equation*}
    \frac{d \mathbb{E}[\rvx{XX}^T]}{dt} = \boldsymbol{\theta}\mathbb{E}[\rvx{XX}^T] + \mathbb{E}[\rvx{XX}^T] \boldsymbol{\theta}^T + \mathbf{B}\mathbb{E}[\rvx{XX}^T]\mathbf{B}^T.
\end{equation*}

Then, forming the covariance matrix yields:
\begin{align*}
    \frac{d\mathbf{P}}{dt} & = \frac{d}{dt} \left( \mathbb{E}[ \rvx{X}\rvx{X}^T ] - \mathbf{mm}^T \right)  \\
    & =  \boldsymbol{\theta} \mathbb{E}[\rvx{XX}^T] +\mathbb{E}[\rvx{XX}^T] \boldsymbol{\theta}^T + \mathbf{B}\mathbb{E}[\rvx{XX}^T]\mathbf{B}^T  - \boldsymbol{\theta}\mathbf{mm}^T - \mathbf{mm}^T \boldsymbol{\theta}^T \\
    & = \boldsymbol{\theta} \left( \mathbb{E}[ \rvx{X}\rvx{X}^T ] - \mathbf{mm}^T \right)  +  \left( \mathbb{E}[ \rvx{X}\rvx{X}^T ] - \mathbf{mm}^T \right) \boldsymbol{\theta}^T  \mathbf{B}\mathbb{E}[ \rvx{X}\rvx{X}^T ]\mathbf{B}^T\\
    & = \boldsymbol{\theta} \mathbf{P} + \mathbf{P}\boldsymbol{\theta}^T  + \mathbf{B}(\mathbf{P} + \mathbf{mm}^T)\mathbf{B}^T,
\end{align*}
which is indeed equal to Eq.~\ref{cov_gbm}. We notice that for the multivariate GBM, the ODEs on the mean and covariance are coupled, since the equation on the covariance involves the mean. In practice, we first integrate the equation on the mean and plug the solution into the covariance ODE.

Finally, choosing $\boldsymbol{\xi}$ as in Eq.~\eqref{xigbm} puts this equation within our integration framework.

\subsubsection{Riccati differential equation in continuous time finite horizon linear quadratic optimal control}

In this last application, we switch to a linear quadratic optimal control problem with finite horizon and continuous time~\cite{bacsar1998dynamic}. We consider a state variable $\mathbf{x} \in \mathbb{R}^n$, initially at $\mathbf{x}(0) = \mathbf{x}_{0}$, that is subject to a linearly controlled linear dynamical system:
\begin{equation*}
    \frac{d\mathbf{x}}{dt} = \mathbf{Ax} + \mathbf{Bu},
\end{equation*}
where $\mathbf{u}\in \mathbb{R}^{n}$ is the control variable, and $\mathbf{A,B}\in \mathbb{R}^{n \times n}$. We want to control the system, e.g. to stabilize the trajectories with $t\in [0,t_f]$, if $\mathbf{A}$ is such that without control, the system diverges. This will happen if $\mathbf{A}$ has at least one eigenvalue with a positive real part. In this case, the goal is then to minimize the quadratic cost function wrt to $\mathbf{u}$, leading to a so called Linear Quadratic Regulator (LQR):
\begin{equation*}
    \mathcal{L}(\mathbf{u}) = \frac{1}{2}  \left(||\mathbf{x}(t_f)||_{\mathbf{Q}(t_f)}^2 + \int_{0}^{t_f} (||\mathbf{x}||^{2}_{\mathbf{Q}} + ||\mathbf{u}||_{\mathbf{R}}^2) dt \right),
\end{equation*}
depending on the choices of the SPD matrices  $\mathbf{Q}_{t_f}$, $\mathbf{Q}$ and $\mathbf{R}$ to set a tradeoff between the energy of the control and how much we want to push the trajectory of $\mathbf{x}$ towards zero. We let $||\mathbf{x}||_{\boldsymbol{\Sigma}}^2 = \mathbf{x}^T\boldsymbol{\Sigma} \mathbf{x}$ for any $\mathbf{x} \in \mathbb{R}^{n}$, and $\boldsymbol{\Sigma}\in \mathrm{Sym}_{n}^{+}$.

The closed-loop control solution is given in closed form by $\mathbf{u}(t) = \mathbf{K}(t)\mathbf{x}(t)$, where $\mathbf{K}(t)$ is a time dependent gain, given by
\begin{equation*}
    \mathbf{K}(t) = \mathbf{R}^{-1}\mathbf{B}^{T}\mathbf{P}(t).
\end{equation*}
In the expression of the gain, $\mathbf{P}(t) \in \mathrm{Sym}_{n}^{+}$ is an SPD matrix that can be obtained by solving the so-called Riccati differential equation:
\begin{equation*}
    \frac{d\mathbf{P}}{dt} = -(\mathbf{AP} + \mathbf{PA}^T - \mathbf{PBR}^{-1}\mathbf{B}^T\mathbf{P} + \mathbf{Q}),
\end{equation*}
with terminal condition $\mathbf{P}(t_f) = \mathbf{Q}_{t_f}$. This time varying algebraic Riccati equation can be derived from the cost function, either using Pontryagin's maximum principle or dynamic programming, through the Hamilton-Jacobi-Bellman equations~\cite{bacsar1998dynamic}. 

Once again, our proposed framework can handle this differential equation (and its many variants) by choosing $\boldsymbol{\xi}(\mathbf{P},t) = -\mathbf{A} + \frac{1}{2} \mathbf{PBR}^{-1}\mathbf{B}^T - \frac{1}{2}\mathbf{QP}^{-1}$.

\subsection{Riemannian metrics on $\mathrm{Sym}_{n}^{+}$}

In this section, we simply give the expression of the Riemannian distance on $\mathrm{Sym}_{n}^{+}$ used in this paper, i.e. the affine invariant distance. We refer to~\cite{arsigny2007geometric},\cite{bhatia2009positive},\cite{bonnabel2010riemannian} for more details on this and other distances on  $\mathrm{Sym}_{n}^{+}$ and the associated Riemannian metrics.

The affine-invariant (sometimes called Fisher-Rao) distance~\cite{bhatia2009positive,bonnabel2010riemannian} is defined as
\begin{equation*}
    d_{A}(\mathbf{P}_1,\mathbf{P}_{2}) = ||\log (\mathbf{P}_1 ^{-1/2} \mathbf{P}_{2} \mathbf{P}_1 ^{-1/2})||_{F}.
\end{equation*}

The corresponding Riemann exponential map $\textrm{Exp}_{\mathbf{P}}^{A}$ at $\mathbf{P} \in \mathrm{Sym}_{n}^{+}$ , used in the Riemannian-RK4 method can be obtained as as~\cite{bhatia2009positive}:
\begin{equation*}
    \textrm{Exp}_{\mathbf{P}}^{A}(\boldsymbol{\Sigma}) =\mathbf{P}^{1/2}\exp(\mathbf{P}^{-1/2}\boldsymbol{\Sigma}\mathbf{P}^{-1/2})\mathbf{P}^{1/2}.
\end{equation*}

%
%

\AtEndDocument{%
\bibliographystyle{splncs04}
\bibliography{refs.bib}
}

%




\end{document}